# Spin-orbit torques acting upon a perpendicularly-magnetized Py layer

Tian-Yue Chen[1†], Yongxi Ou[3†], Tsung-Yu Tsai[1], R. A. Buhrman[3], and Chi-Feng Pai[1,2*]

[1]*Department of Materials Science and Engineering, National Taiwan University, Taipei 10617, Taiwan*

[2]*Center of Atomic Initiative for New Materials, National Taiwan University, Taipei 10617, Taiwan*

[3]*Cornell University, Ithaca, New York 14853, USA*

We show that Py, a commonly-used soft ferromagnetic material with weak anisotropy, can become perpendicularly-magnetized while depositing on Ta buffer layer with Hf or Zr insertion layers (ILs) and MgO capping layer. By using two different approaches, namely harmonic voltage measurement and hysteresis loop shift measurement, the dampinglike spin-orbit torque (DL-SOT) efficiencies from Ta/IL/Py/IL/MgO magnetic heterostructures with perpendicular magnetic anisotropy are characterized. We find that though Ta has a significant spin Hall effect, the DL-SOT efficiencies are small in systems with the Ta/Py interface compared to that obtained from the control sample with the traditional Ta/CoFeB interface. Our results indicate that the spin transparency for the Ta/Py interface is much less than that for the Ta/CoFeB interface, which might be related to the variation of spin mixing conductance for different interfaces.

---

[†] These authors contributed equally to this work.
[*] Email: cfpai@ntu.edu.tw

Current-induced spin-orbit torque (SOT) is an efficient mechanism to drive magnetic dynamics,[1-3] domain wall motion,[4,5] as well as magnetization switching[6-8] in various magnetic heterostructures. These heterostructures typically consist of a heavy metal (HM)/ferromagnetic metal (FM) bilayer or a HM/FM/oxide trilayer design. When a charge current is flowing in the HM layer with large spin-orbit interaction, such as Ta, W, or Pt, the spin Hall effect (SHE)[9] therein generates a transverse spin current that will flow into the FM layer. However, the transmission of this spin current or spin transparency at the HM/FM interface will also strongly affect the resulting SOT efficiency acting on the FM layer, as can be seen in various Pt/FM heterostructures.[10,11] The effective dampinglike SOT (DL-SOT) efficiency of a particular heterostructure therefore should be expressed as $\xi_{\text{DL}}^{\text{HM/FM}} = T_{\text{int}}^{\text{HM/FM}} \theta_{SH}^{\text{HM}}$, where $\theta_{SH}^{\text{HM}}$ and $T_{\text{int}}^{\text{HM/FM}}$ represent the internal spin Hall ratio of HM and the HM/FM interfacial spin transparency factor, respectively.

Various characterization techniques have been developed to estimate $\xi_{\text{DL}}^{\text{HM/FM}}$, such as spin-torque ferromagnetic resonance (ST-FMR),[2] harmonic voltage[12] or Kerr rotation[13] measurement, chiral domain wall motion detection,[14] and current-induced loop shift measurement.[15] Inconveniently, the above-mentioned techniques do not apply to all kinds of magnetic heterostructures. For instance, if the FM layer in the heterostructure has perpendicular magnetic anisotropy (PMA), harmonic measurement and loop shift approach can be employed. The current-induced SOT switching can also be demonstrated in micron-sized Hall-bar devices without the need of nano-sized device patterning. In contrast, while dealing with magnetic heterostructures with in-plane magnetized FM layer such as Py (Ni$_{80}$Fe$_{20}$), ST-FMR measurement is commonly used to estimate $\xi_{DL}^{HM/Py}$.[2] Therefore, the options of SHE and SOT characterization techniques for HM/Py systems are mainly limited by the magnetic anisotropy nature of Py layer. Recently, it was reported by Ou *et. al.* that the Py layer in an as-grown Ta/Py/HfO$_2$/MgO structure (HfO$_2$ is naturally oxidized from Hf by the adjacent MgO layer) can possess PMA with a sizable anisotropy field of $H_{an}$ ~ 1000 Oe,[16] which motivates us to further utilize this heterostructure to characterize SOT efficiencies in systems with perpendicularly-magnetized Py.

In this work, we show that by introducing suitable insertion layers (ILs), the weak-anisotropy Py layer can obtain PMA in Ta-based magnetic heterostructures. We further characterize the SOT properties of such layer structures by both harmonic voltage measurement[12] and current-induced hysteresis loop shift measurement.[15] We find that the choice of ferromagnetic layer will drastically change the resulting dampinglike SOT (DL-SOT) efficiency. Unlike the conventional Ta-based heterostructure with the Ta/CoFeB interface, which has $\xi_{DL}^{Ta/CoFeB} \approx -0.09$ from our control sample, the heterostructures with the Ta/Py interface only possess a much weaker $\xi_{DL}^{Ta/Py} \approx -0.03$ (harmonic method) or $\xi_{DL}^{Ta/Py} \approx -0.01$ (loop shift method). This result is an indication of different spin transparency[10, 11] between the Ta/CoFeB and the Ta/Py interfaces. Nevertheless, even if $\left|\xi_{\text{DL}}^{\text{Ta/Py}}\right|$ is at least three times smaller than $\left|\xi_{\text{DL}}^{\text{Ta/CoFeB}}\right|$, current-induced SOT switching can still be achieved in Ta/Py PMA devices. The Ta/Py PMA heterostructure therefore is suitable for SOT studies on weak anisotropy magnetic layers.

The Ta/IL/Py/IL/MgO heterostructures studied in this work were deposited via standard direct current (DC) sputtering (with RF magnetron sputtering for the MgO layer) onto Si/SiO$_2$ substrates. The base pressure was kept $<4\times10^{-8}\,\text{Torr}$, with the DC sputtering condition of 2mTorr Ar pressure and 30 watts power. Two kinds of materials, namely Hf and Zr, were selected as the ultrathin ILs with optimized thickness ($\leq 0.5\,\text{nm}$) to attain the PMA in the Py layer. The Hf and Zr ILs were sputtered with a low deposition rate of $0.01\,\text{nm/s}$. We note that the thin ILs between the Py and MgO layers became fully oxidized due to the deposition of MgO on top of them,[16] which protects the Fe atoms in FM from over-oxidation, therefore enhancing the PMA. The Fe-O-Hf(Zr) bonds may also provide stronger spin-orbit splitting of the Fe-O orbitals at the Fermi level and further enhance the interfacial PMA effect.

To examine the perpendicular-magnetized Py layer microstructure, we first performed cross-sectional field-emission transmission electron microscopy (FE-TEM, JEOL 2010F) on Ta(6)/Py(1.5)/Hf(0.2)/MgO(2)/Ta(1.5) and Ta(6)/Zr(0.5)/Py(1.8)/Zr(0.2)/MgO(2)/Ta(1.5) films.

The TEM samples were prepared by a lift-out technique with SEIKO SMI-3050SE focused ion beam (FIB). As shown in Fig.1(a) and (b), although the deposited layers are fairly uniform, the sputtered Py layer and the buffer layers have no observable crystalline structures. The out-of-plane hysteresis loops obtained from magneto-optical Kerr effect (MOKE) further indicate that both films with Hf and Zr ILs have PMA, as shown in Fig.1(c) and (d).

We first present the magnetic properties and SOT properties from Ta/Py/Hf/MgO samples. The magnetic properties of Ta/Py/Hf/MgO heterostructures were characterized by vibrating sample magnetometer (VSM). The effective saturation magnetization of Py in the Ta/Py/Hf/MgO system is found to be $M_s$ = 1079 emu/cm$^3$ with a magnetic dead layer of $t_{dead}$ ~ 0.9 nm, as shown in Fig. 2(a). It is noted that because of the relatively thick dead layer in the Ta/Py/Hf/MgO heterostructure, the effective $M_s$ is greater than the bulk value.[17] Fig. 2(b) shows the effective anisotropy energy densities for a series of Ta/Py/Hf/MgO samples. The $K_{eff} \cdot t_{Py}^{eff}$ vs $t_{Py}^{eff}$ plot indicates that the perpendicular magnetized Py exists in the thickness range of 0.5 nm < $t_{Py}^{eff}$ < 0.7 nm.

To perform current-induced magnetization switching measurements and detailed SOT characterizations, we patterned a representative film, Ta(6)/Py(1.5)/Hf(0.2)/MgO(2) into Hall bar devices with lateral dimension 5 μm×60 μm via standard photolithography processes, as shown in Fig. 2(c). The switching measurement was done by sending current pulses $I_{sw}$ of 100 ms duration into the device while applying an in-plane bias field. The purpose of the external in-plane field is to break the inversion symmetry and/or realign the chiral domain wall moments.[18] The spin current generated from the SHE of the Ta buffer layer can exert DL-SOT on the contiguous FM layer, which can further switch the magnetization direction therein. As shown in Fig. 2(d), current-induced switching of the Ta(6)/Py(1.5)/Hf(0.2)/MgO(2) PMA device can be realized by $I_{sw} \approx \pm 2\,\text{mA}$. The switching polarity will change according to the direction of $H_x$, which is consistent with the conventional DL-SOT switching scenario in other Ta-based devices.[7] The steps in the switching loop can be considered as the evidence of current-driven domain wall motion in these micron-sized devices.

In order to determine the strength of the SOTs in our perpendicularly magnetized Py samples, we first performed the harmonic voltage measurements on a Ta(6)/Py(1.4)/Hf(0.2)/MgO(2) Hall-bar device, with the measurement schematics shown in Fig. 3(a). An AC current ($f \sim 1000$ Hz) was applied along the $x$ direction of the sample while the sample was also kept under a sweeping magnetic field along either the longitudinal ($H_x$) or transverse ($H_y$) direction. The first and second harmonic anomalous Hall (AH) voltages were measured from the transverse Hall channel. As shown in Fig. 3(c), the two parabolic branches in the first harmonic AH voltage $V_\omega$ of the sample are consistent with Py exhibiting PMA. The second harmonic AH voltages $V_{2\omega}$ show linear magnetic field dependence for both $H_x$ and $H_y$ field sweep, as can be seen in Fig. 3(d). We can evaluate the DL and fieldlike (FL) SOT-induced effective fields $\Delta H_{DL(FL)}$ from the first and second harmonic voltages:[12,19]

$$\Delta H_{x(y)} = -2\frac{\partial V_{2\omega}/\partial H_{x(y)}}{\partial^2 V_\omega/\partial H_{x(y)}^2} \qquad (1)$$

$$\Delta H_{DL(FL)} = \frac{\Delta H_{x(y)} + 2\delta \cdot \Delta H_{y(x)}}{1-4\delta^2}, \qquad (2)$$

where $\delta = 0.08$ is the ratio of the planar Hall voltage to the AH voltage. The SOT efficiencies can then be calculated via the effective fields per current density in the Ta layer $\Delta J$:[20]

$$\xi_{DL(FL)} = \frac{2e}{\hbar}\mu_0 M_s \left(t_{Py} - t_{dead}\right)\frac{\Delta H_{DL(FL)}}{\Delta J}. \qquad (3)$$

Based on above equations, we determined the DL-SOT and FL-SOT efficiencies for the Ta(6)/Py(1.4)/Hf(0.2)/MgO(2) sample as $\xi_{DL}^{Ta/Py} = -0.03 \pm 0.01$ and $\xi_{FL}^{Ta/Py} = 0.13 \pm 0.01$, respectively.

For further characterizing the DL-SOT efficiency of Ta/Py/Hf/MgO heterostructures that is more pertinent to the nature of magnetization switching, we also performed current-induced hysteresis loop shift measurement on the above-mentioned patterned Hall-bar devices, as shown in Fig. 3(b). While injecting DC currents into the Hall-bar devices, we again have to apply an in-plane field in order to break the symmetry and/or realign the domain wall moments thus the DL-SOT generated from the buffer layer can act on the magnetic moment in the Py layer. Representative results from a Ta(6)/Py(1.5)/Hf(0.2)/MgO(2) device are shown in Fig. 4(a), in which the DL-SOT-generated out-of-plane effective field $H_z^{eff}$ can be detected through the shift of AH loops. Fig. 4(b) shows that the Ta/Py/Hf/MgO heterostructure has good thermal stability since the current-induced joule heating has little effect on the coercive field of the Ta/Py heterostructure. The linear trend of $H_z^{eff}$ (center of the hysteresis loop) vs. $I_{dc}$ again indicates the existence of a current-induced out-of-plane effective field stemming from DL-SOT. In Fig. 4(c), the summarized $H_z^{eff}$ as functions of applied current $I_{dc}$ under opposite in-plane fields are shown. The opposite linear trends for opposite $H_x$ again satisfies the DL-SOT-driven switching scenario, which originates from the opposite alignment of domain-wall moment in the Py layer.

The current-induced effective field per current density is typically expressed as $\chi \equiv H_z^{eff}/J_e$,[14] where $J_e$ is the charge current density flowing in the Ta buffer layer. In our Ta/Py/Hf/MgO heterostructures, the measured $\chi$ saturates at $\chi_{sat} \approx -5\,\text{Oe}/10^{11}\,\text{A}\cdot\text{m}^{-2}$ when the applied in-plane field is greater than $H_x \approx 200\,\text{Oe}$ because the field is large enough to overcome Dzyaloshinskii-Moriya interaction (DMI) and to fully realign the domain wall moments.[4] This DMI effective field translates to a DMI constant of $D \sim 0.5$ mJ/m$^2$, which is smaller than the typical reported values of $D \sim 1.0$ mJ/m$^2$ for the Pt/Py interface.[21, 22] Note that the magnitude of $\chi_{sat}$ in Ta/Py system is smaller than that of Ta/CoFeB systems.[15] To further extract the DL-SOT efficiency from $\chi_{sat}$, again we use:[20]

$$\xi_{DL} = \frac{2e}{\hbar}\left(\frac{2}{\pi}\right)\mu_0 M_s \left(t_{Py} - t_{dead}\right)\chi_{sat}, \qquad (4)$$

where $M_s$ is the saturation magnetization determined by VSM. The DL-SOT efficiency is therefore estimated to be $\xi_{DL}^{Ta/Py} \approx -0.01$ from the loop shift measurement.

We also characterized the SOTs from a Ta(6)/Zr(0.5)/Py(1.5)/Zr(0.2)/MgO(2) sample that has Zr ILs on both sides of the Py to enhance its PMA (Ou *et al.* unpublished). The first and second harmonic AH voltages are shown in Fig. 5(a) and (b) respectively. With Eq.(1)-(3), we can estimate the DL and FL-SOT efficiencies of this sample as $\xi_{DL}^{Ta/Py} = -0.05 \pm 0.02$ and $\xi_{FL}^{Ta/Py} = 0.06 \pm 0.02$, respectively. As shown in Fig. 5(c) and (d), the hysteresis loop shift approach also suggests a small DL-SOT efficiency $\xi_{DL}^{Ta/Py} \approx -0.01$ in the Ta/IL/Py/IL system. To compare, we determined DL-SOT efficiency to be $\xi_{DL}^{Ta/Py} \approx -0.09$ from a Ta/CoFeB/Hf/MgO control sample with PMA.

The small $|\xi_{DL}^{Ta/Py}| \approx 0.01$ of Ta/Py PMA system while compare to $|\xi_{DL}^{Ta/CoFeB}| \approx 0.09$ and other larger reported values of $\left|\xi_{\text{DL}}^{\text{Ta/CoFeB}}\right| \approx 0.10$ [7, 15] in Ta/CoFeB PMA system is perhaps due to the poor band matching between Py and Ta layers.[23] A similar trend has also been reported by Kondou *et. al.* for the in-plane magnetized case, in which they found $\left|\xi_{\text{DL}}^{\text{Ta/Py}}\right| \approx 0.01$ and $\left|\xi_{\text{DL}}^{\text{Ta/CoFeB}}\right| \approx 0.03$ via the ST-FMR approach with inverse SHE (ISHE) voltage correction[24] and they attributed the difference between Ta/Py and Ta/CoFeB systems to the spin resistance mismatch.[25]

In summary, we first show that the soft magnetic material Py can possess perpendicular anisotropy under suitable combination of insertion layers. Second, even though the DL-SOT efficiency in the Ta/Py system is smaller than that in the Ta/CoFeB system, current-induced switching can be demonstrated. The DL-SOT efficiency from Ta/Py heterostructures, as determined by two different characterization techniques, is only $\left|\xi_{\text{DL}}^{\text{Ta/Py}}\right| \approx 0.01-0.03$ due to the possible poor band matching between Ta and Py layers. The DMI constant of the Ta/Py interface is $D \sim 0.5$ mJ/m$^2$, which is smaller than the reported value of $D \sim 1.0$ mJ/m$^2$ for the Pt/Py interface. Our results suggest that the perpendicular magnetized HM/Py system can be used to study SOT-related phenomenon beyond conventional HM/CoFeB heterostructures. If one can further introduce perpendicularly-magnetized Py on Pt layer, then such PMA materials system can be possibly extended to study chiral domain wall motion as well as skyrmion dynamics due to the large interfacial DMI at the Pt/Py interface.

**Acknowledgments**

This work was partly supported by the Ministry of Science and Technology of Taiwan under grant No. MOST 105-2112-M-002-007-MY3 and by the Center of Atomic Initiative for New Materials (AI-Mat), National Taiwan University, from the Featured Areas Research Center Program within the framework of the Higher Education Sprout Project by the Ministry of Education (MOE) in Taiwan under grant No. NTU-107L9008. This work was also partly supported by ONR (N000014-15-1-2449) and by NSF/MRSEC (DMR-1120296) through the Cornell Center for Materials Research (CCMR), and by NSF through use of the Cornell NanoScale Facility, an NNCI member (ECCS-1542081).

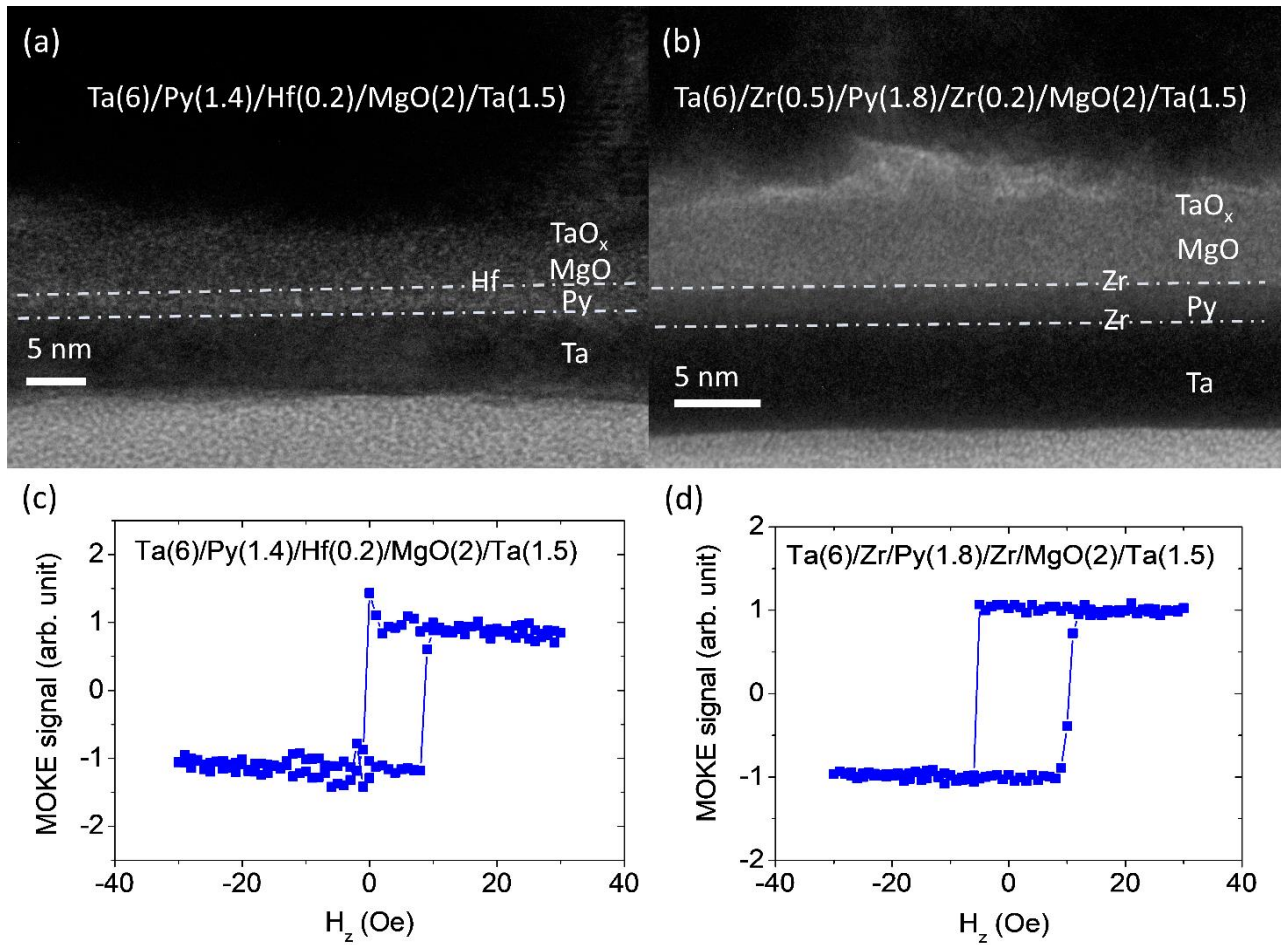

Figure 1. Cross-sectional TEM images of sputtered (a) Ta(6)/Py(1.4)/Hf(0.2)/MgO(2) and (b) Ta(6)/Zr(0.5)/Py(1.8)/Zr(0.2)/MgO(2) PMA films. The dash-dotted lines are guidelines to indicate the interfaces. Out-of-plane hysteresis loops of (c) Ta/Py/Hf/MgO and (d) Ta/Zr/Py/Zr/MgO films obtained from MOKE measurements. The shift of hysteresis loops towards positive $H_z$ is due to the remnant field (~ 5 Oe) from the pole piece of electromagnet.

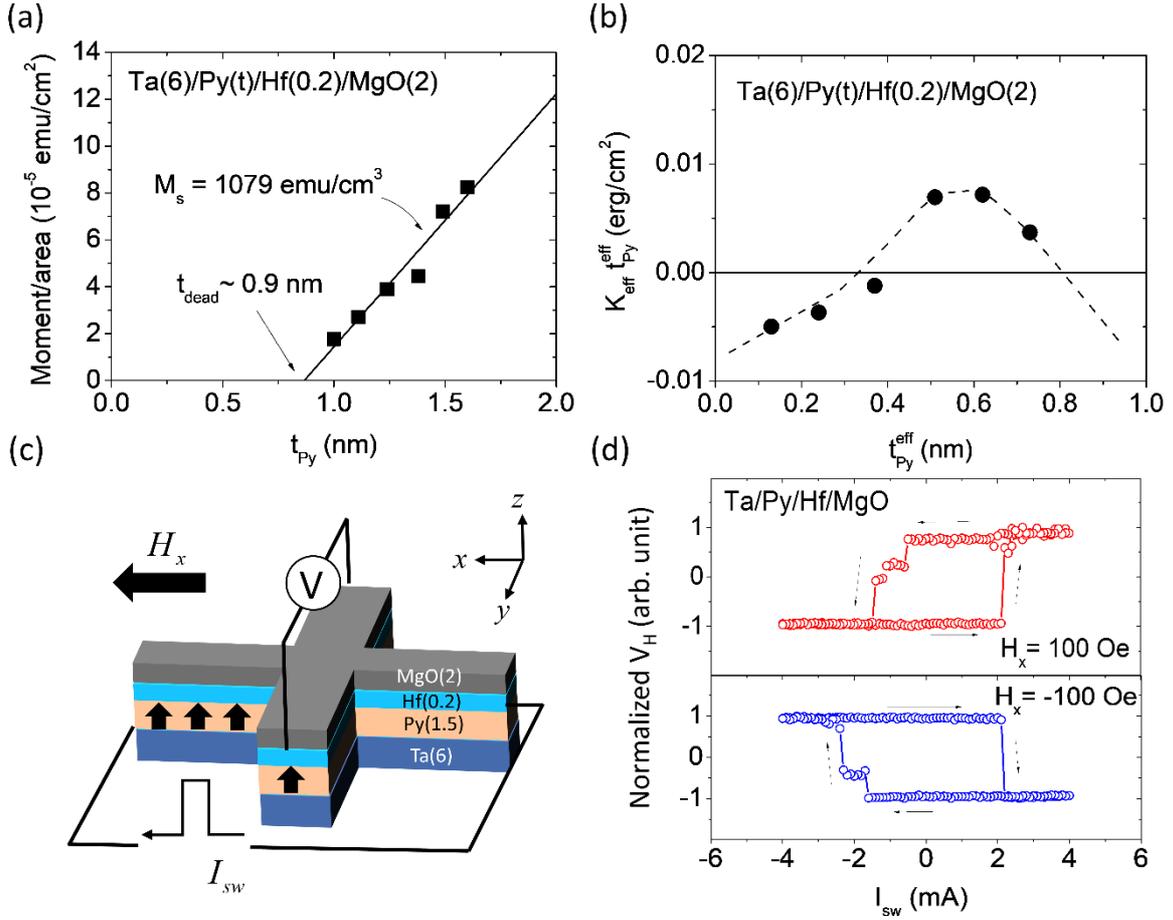

Figure 2. (a) Moment per area of Ta(6)/Py($t_{Py}$)/Hf(0.2)/MgO(2) films as a function of Py thickness $t_{Py}$. (b) Effective anisotropy energy density $K_{eff}$-effective thickness $t_{Py}^{eff}$ product of Ta(6)/Py($t_{Py}$)/Hf(0.2)/MgO(2) films as a function of effective Py thickness $t_{Py}^{eff} = t_{Py} - t_{dead}$. (c) Schematic illustration for current-induced switching measurement of a micron-sized Ta(6)/Py(1.5)/Hf(0.2)/MgO(2) Hall-bar sample. $I_{sw}$ represents the amplitude of applied current pulse. $H_x$ is the applied in-plane bias field. (d) Current-induced SOT switching loops of a Ta/Py/Hf/MgO Hall-bar device measured with $H_x = \pm 100$ Oe.

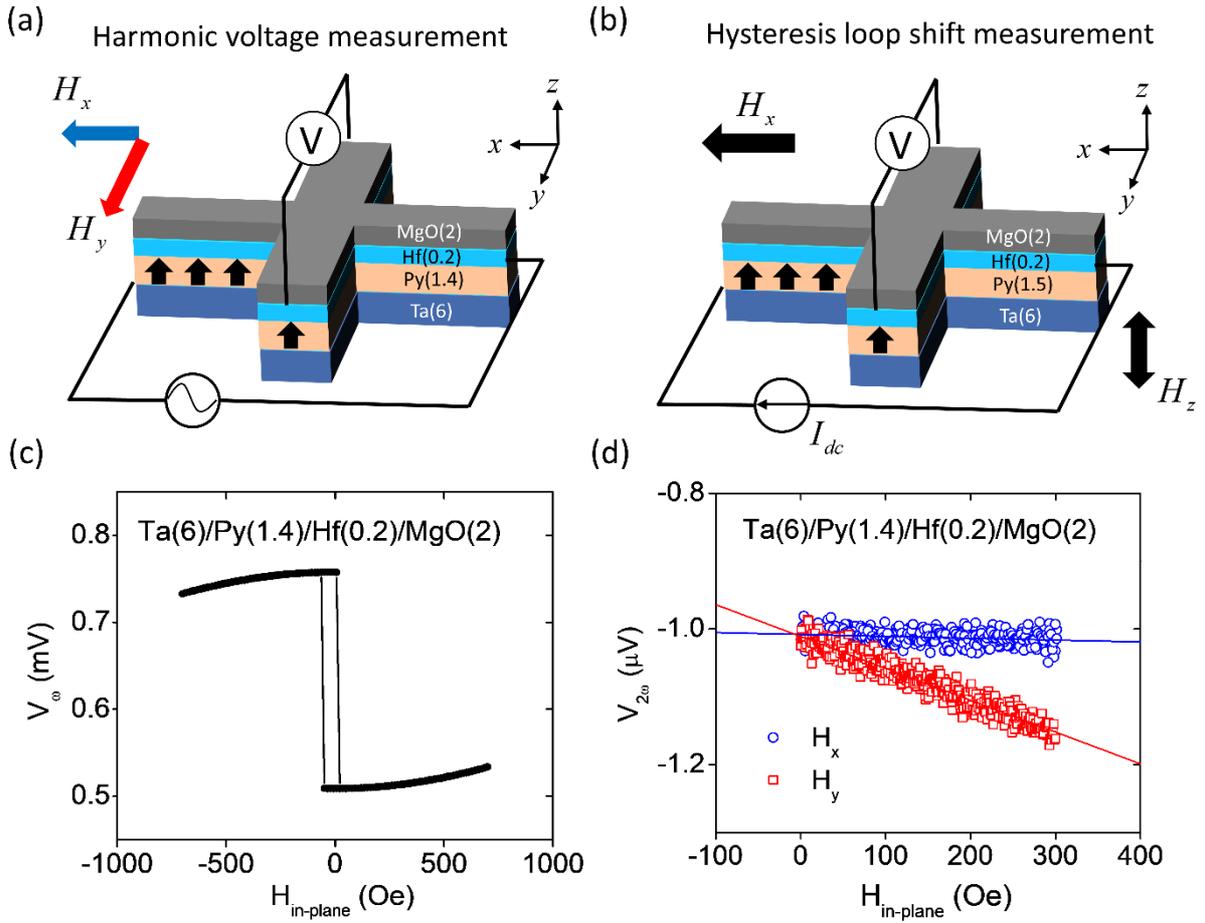

Figure 3. Schematic illustrations of (a) harmonic voltage measurement and (b) hysteresis loop shift measurement. Representative (c) first harmonic and (d) second harmonic voltage measurement results from a Ta(6)/Py(1.4)/Hf(0.2)/MgO(2) Hall-bar device. The blue circles and red squares in (d) represent second harmonic data obtained with longitudinal in-plane filed $H_x$ scan and transverse in-plane filed $H_y$ scan, respectively.

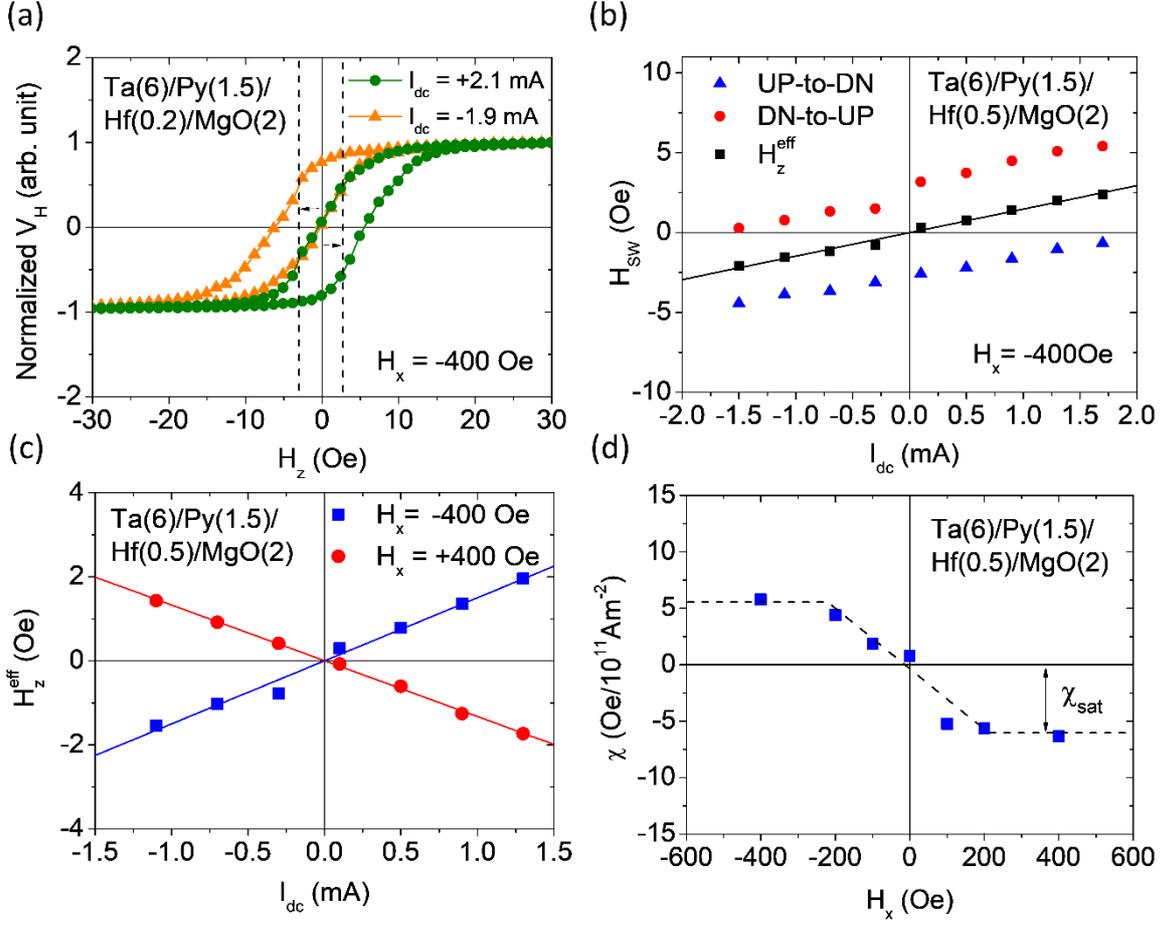

Figure 4. (a) Current-induced hysteresis loop shift measurement results obtained from a Ta/Py/Hf/MgO Hall-bar sample with $H_x = -400\,\mathrm{Oe}$ and different applied currents $I_{dc}$. (b) The out-of-plane switching fields $H_{sw}$ (red circles and blue triangles) of a Ta/Py/Hf/MgO Hall-bar sample as functions of $I_{dc}$. The black squares represent the center of measured hysteresis loops, $H_z^{eff}$. (c) Current-induced effective field $H_z^{eff}$ of Ta/Py/Hf/MgO as functions of $I_{dc}$ for $H_x = \pm 400\,\mathrm{Oe}$. (d) Current-induced effective field per current density $\chi$ of Ta/Py/Hf/MgO as a function in-plane bias field $H_x$.

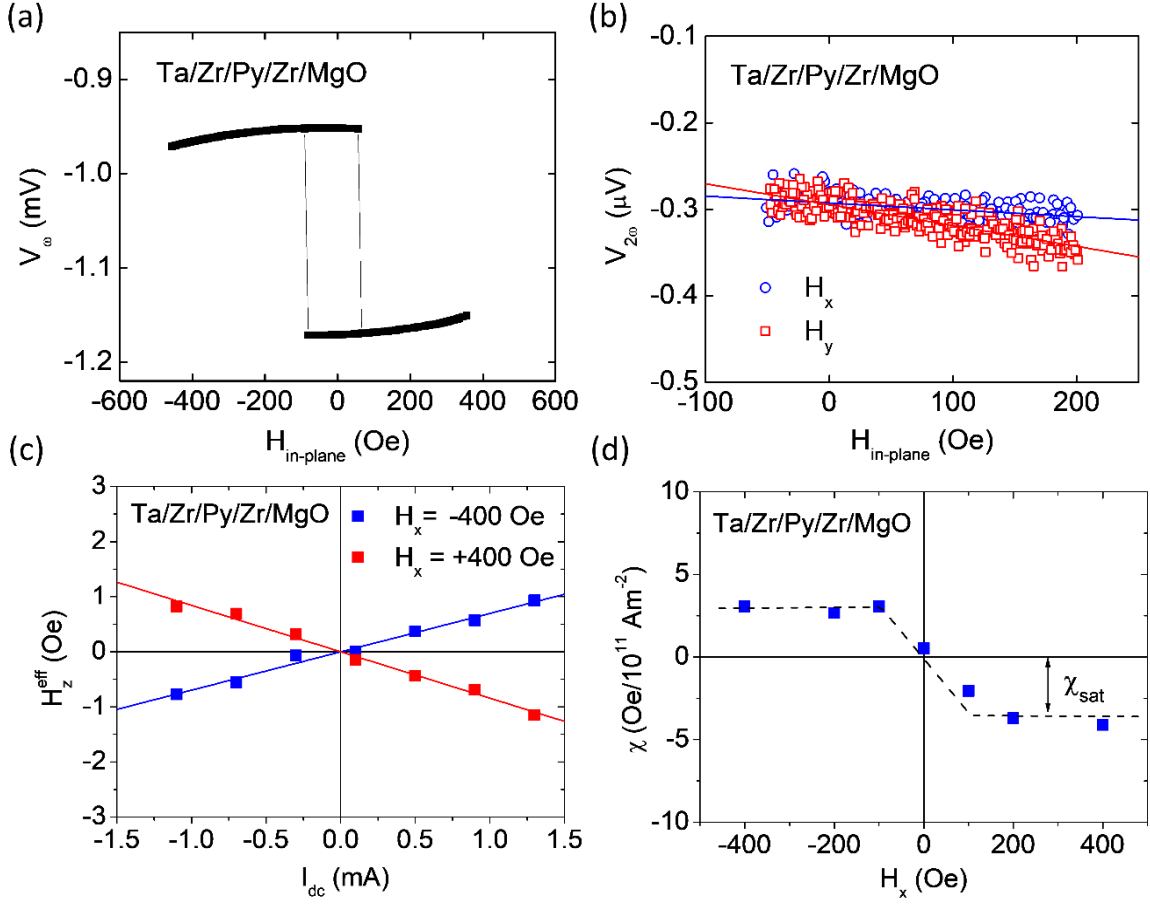

Figure 5. Representative (a) first harmonic and (b) second harmonic voltage measurement results from a Ta(6)/Zr(0.5)/Py(1.5)/Zr(0.2)/MgO(2) Hall-bar device. (c) Current-induced effective field $H_z^{eff}$ of a Ta/Zr/Py/Zr/MgO Hall-bar device as functions of $I_{dc}$ for $H_x = \pm 400\,\text{Oe}$. (d) Current-induced effective field per current density $\chi$ of Ta/Zr/Py/Zr/MgO as a function in-plane bias field $H_x$.